\documentclass[nonacm,sigconf,natbib=true]{acmart}
\usepackage[norelsize,ruled,vlined,linesnumbered]{algorithm2e}
\usepackage{algorithmic}
\SetKwComment{Comment}{$\triangleright$\ }{}
\usepackage{subfigure}
\usepackage{booktabs}
\usepackage{cleveref}

\AtBeginDocument{%
  }

\newtheoremstyle{mystyle}
    {1.0mm}
    {1.0mm}
    {\it}
    {0.0mm}
    {\scshape}
    {.}
    { }
    {}
\theoremstyle{mystyle}
\newtheorem{definition}{Definition}

\newtheorem{lemma}{Lemma}

\newcommand{\vs}{\vspace{1.0mm}}

\setcopyright{acmlicensed}
\copyrightyear{2018}
\acmYear{2018}
\acmDOI{XXXXXXX.XXXXXXX}

\setcopyright{none}
\renewcommand\footnotetextcopyrightpermission[1]{} 
\acmConference[Conference acronym 'XX]{Make sure to enter the correct conference title from your rights confirmation emai}{June 03--05, 2018}{Woodstock, NY}
\acmISBN{978-1-4503-XXXX-X/18/06}

\begin{document}

\title{Approximate Reverse $k$-Ranks Queries in High Dimensions}

\author{Daichi Amagata}
\authornote{Both authors contributed equally to this research. Daichi Amagata is the corresponding author.}
\email{amagata.daichi@ist.osaka-u.ac.jp}
\author{Kazuyoshi Aoyama}
\authornotemark[1]
\email{aoyama.kazuyoshi@ist.osaka-u.ac.jp}
\affiliation{%
  \institution{The University of Osaka}
  \city{Suita, Osaka}
  \country{Japan}
}
\author{Keito Kido}
\email{kido.keito@ist.osaka-u.ac.jp}
\affiliation{%
  \institution{The University of Osaka}
  \city{Suita, Osaka}
  \country{Japan}
}
\author{Sumio Fujita}
\email{sufujita@lycorp.co.jp}
\affiliation{%
  \institution{LY Corporation}
  \city{Chiyoda, Tokyo}
  \country{Japan}
}

\renewcommand{\shortauthors}{Aoyama et al.}

\begin{abstract}
Many objects are represented as high-dimensional vectors nowadays.
In this setting, the relevance between two objects (vectors) is usually evaluated by their inner product.
Recently, item-centric searches, which search for users relevant to query items, have received attention and find important applications, such as product promotion and market analysis.
To support these applications, this paper considers reverse $k$-ranks queries.
Given a query vector $\mathbf{q}$, $k$, a set $\mathbf{U}$ of user vectors, and a set $\mathbf{P}$ of item vectors, this query retrieves the $k$ user vectors $\mathbf{u} \in \mathbf{U}$ with the highest $r(\mathbf{q},\mathbf{u},\mathbf{P})$, where $r(\mathbf{q},\mathbf{u},\mathbf{P})$ shows the rank of $\mathbf{q}$ for $\mathbf{u}$ among $\mathbf{P}$.
Because efficiently computing the exact answer for this query is difficult in high dimensions, we address the problem of approximate reverse $k$-ranks queries.
Informally, given an approximation factor $c$, this problem allows, as an output, a user $\mathbf{u}'$ such that $r(\mathbf{q},\mathbf{u}',\mathbf{P}) > \tau$ but $r(\mathbf{q},\mathbf{u}',\mathbf{P}) \leq c \times \tau$, where $\tau$ is the rank threshold for the exact answer.
We propose a new algorithm for solving this problem efficiently.
Through theoretical and empirical analyses, we confirm the efficiency and effectiveness of our algorithm.
\end{abstract}





\maketitle

\section{Introduction}
Due to the machine learning-based embedding techniques, many objects have been represented as high-dimensional vectors.
In this setting, the relevance between two objects (vectors) is usually evaluated by their inner product.
Therefore, inner product retrieval has been studied in many fields, such as information retrieval \cite{aoyama2023simpler}, data mining \cite{tan2021norm}, databases \cite{nakama2021approximate}, artificial intelligence \cite{zhang2023query}, machine learning \cite{guo2020accelerating}, recommender systems \cite{hirata2022solving}, and others \cite{hirata2022cardinality,hirata2023categorical}.

\vs
\noindent
\textbf{Motivation and challenge.}
The above works can be classified into two searches: user-centric searches and item-centric searches.
Note that we use $\mathbf{U}$ and $\mathbf{P}$ to denote sets of user and item vectors, respectively.

The $k$-MIPS (Maximum Inner Product Search) problem \cite{morozov2018non} represents the user-centric search problem.
Given a user vector $\mathbf{u} \in \mathbf{U}$ and $\mathbf{P}$, this problem finds the item vector $\mathbf{p^{*}} =$ arg$\,\max_{\mathbf{p}\in \mathbf{P}}\mathbf{p} \cdot \mathbf{u}$.
That is, its output is the item most relevant to the given user.

Reverse $k$-MIPS \cite{amagata2021reverse,amagata2023reverse} and reverse $k$-ranks \cite{bian2024qsrp} queries are two representative item-centric searches.
Informally, these queries find users interested in a given query item vector.
The reverse $k$-MIPS problem is a reverse version of the $k$-MIPS problem.
Given an item vector $\mathbf{p} \in \mathbf{P}$ and $\mathbf{U}$, this problem finds all user vectors in $\mathbf{U}$ such that their $k$-MIPS results contain $\mathbf{p}$.
If a query item vector is popular, the reverse $k$-MIPS query tends to return many users.
However, if a query item vector is niche, the query result contains few users.
The reverse $k$-ranks problem removes this size-uncontrollable issue.
Given a query item vector $\mathbf{q}$, a user vector $\mathbf{u} \in \mathbf{U}$, and $\mathbf{P}$, we define $r(\mathbf{q},\mathbf{u},\mathbf{P})$, the rank of $\mathbf{q}$ for $\mathbf{u}$ among $\mathbf{P}$.
Then, this problem finds the $k$ user vectors in $\mathbf{U}$ with the highest $r(\mathbf{q},\cdot,\mathbf{P})$.
The reverse $k$-ranks problem hence guarantees a $k$-sized result.
Due to this advantage, this problem can be used in important applications, including target advertisements and potential customer discovery.
Furthermore, literature \cite{bian2024qsrp} demonstrates that, compared with random, UserPop \cite{he2017neural}, $k$-MIPS, and reverse $k$-MIPS, the reverse $k$-ranks problem yields a higher accuracy in item-centric recommendation.
This paper therefore considers reverse $k$-ranks queries.

A straightforward algorithm for the reverse $k$-ranks problem is to compute $r(\mathbf{q},\mathbf{u},\mathbf{P})$ for every $\mathbf{u} \in \mathbf{U}$.
In high dimensions, a linear scan of $\mathbf{P}$ is the fastest way of exactly obtaining $r(\mathbf{q},\mathbf{u},\mathbf{P})$.
Hence, this algorithm requires $O(nmd)$ time, where $n = |\mathbf{U}|$, $m = |\mathbf{P}|$, and $d$ is the dimensionality.
Because this time is prohibitive when $n$ and $m$ are large, literature \cite{bian2024qsrp} proposes QSRP, an algorithm that alleviates this issue.
Its idea is to summarize the inner products of $\mathbf{U} \times \mathbf{P}$, and it computes the inner products of all pairs of user and item vectors in a pre-processing step.
As can be seen from this bruteforce manner, this approach does not solve the issue held by the straightforward algorithm, and the pre-processing step costs ${\rm\Omega}(nmd)$ time, which is still too large even for pre-processing.
Moreover, its online algorithm does not reduce the theoretical time, i.e., its time complexity is still $O(nmd)$.

\vs
\noindent
\textbf{Contribution.}
To address the above challenge, we make the following contributions.

\vs
\noindent
$\triangleright$ \textit{$c$-approximate reverse $k$-ranks query.}
We first relax the reverse $k$-ranks problem.
Specifically, we propose its variant, i.e., the $c$-approximate reverse $k$-ranks problem.
Let $r(\mathbf{q},\mathbf{u}_{i},\mathbf{P})$ be the $i$-th result in a reverse $k$-ranks query.
The $c$-approximate reverse $k$-ranks problem allows a user vector $\mathbf{u}$ such that $r(\mathbf{q},\mathbf{u},\mathbf{P}) \leq c \times r(\mathbf{q},\mathbf{u}_{i},\mathbf{P})$ as an output, where $c \geq 1$.
That is, the reverse $k$-ranks problem is a special case of our problem (i.e., $c = 1$).
Note that, as represented by existing approximation problems (e.g., the approximate nearest neighbor search problem), many applications require a fast response while allowing approximate results.
This paper solves the $c$-approximate reverse $k$-ranks problem efficiently.

\vs
\noindent
$\triangleright$ \textit{A new algorithm.}
Efficiently solving the $c$-approximate reverse $k$-ranks problem is not trivial, and a straightforward algorithm for this problem remains the same one introduced earlier.
(QSRP \cite{bian2024qsrp} also requires $O(nmd)$ time in this problem.)
To overcome this inefficiency issue, we propose a new approximation algorithm with faster offline and online time than QSRP, and the time of our query processing algorithm does not have a factor of $m$.
In practice, this algorithm quickly returns a highly accurate result against the one for the reverse $k$-ranks problem.

\vs
\noindent
$\triangleright$ \textit{Extensive experiments.}
We conduct experiments on real-world datasets and compare our algorithm with QSRP.
The experimental results demonstrate that our algorithm is much faster than QSRP and yields a competitive accuracy with QSRP.

\section{Preliminary}   \label{sec:preliminary}
Let $\mathbf{U}$ be a set of $n$ user vectors.
Also, let $\mathbf{P}$ be a set of $m$ item vectors.
Each vector in $\mathbf{U}$ and $\mathbf{P}$ is a $d$-dimensional vector, and we assume that $d$ is large (i.e., each vector is a high-dimensional vector).
We use $\mathbf{u} \cdot \mathbf{p}$ to denote the inner product of $\mathbf{u}$ and $\mathbf{p}$.


\begin{definition}[Rank of $\mathbf{q}$ for $\mathbf{u}$]   \label{def:rank}
Given a user $\mathbf{u}$, an item $\mathbf{q}$, and $\mathbf{P}$, the rank of $\mathbf{q}$ for a user $\mathbf{u}$ among $\mathbf{P}$, $r(\mathbf{q},\mathbf{u},\mathbf{P})$, is
\begin{equation*}
    r(\mathbf{q},\mathbf{u},\mathbf{P}) = 1 + \sum_{\mathbf{p} \in \mathbf{P}}\mathbb{I}[\mathbf{u}\cdot \mathbf{p} > \mathbf{u} \cdot \mathbf{q}],
\end{equation*}
where $\mathbb{I}[x]$ is an indicator function and returns 1 (0) if $x$ is true (false).
\end{definition}

\noindent
If $r(\mathbf{q},\mathbf{u},\mathbf{P})$ is small, $\mathbf{u}$ has less item vectors such that $\mathbf{u}\cdot \mathbf{p} > \mathbf{u} \cdot \mathbf{q}$ and relatively prefers to $\mathbf{q}$.
In this concept, reverse $k$-ranks query is defined as follows:

\begin{definition}[Reverse $k$-ranks query]    \label{def:exact}
Given $\mathbf{U}$, $\mathbf{P}$, a query vector $\mathbf{q}$, and $k$, this query returns $\mathbf{U}_{rr} \subseteq \mathbf{U}$ such that $|\mathbf{U}_{rr}| = k$ and for each $\mathbf{u} \in \mathbf{U}_{rr}$ and $\mathbf{u}' \in \mathbf{U}\backslash\mathbf{U}_{rr}$, $r(\mathbf{q},\mathbf{u},\mathbf{P}) \leq r(\mathbf{q},\mathbf{u}',\mathbf{P})$.
\end{definition}

\noindent
Based on this definition, we propose its $c$-approximate version.

\begin{definition}[$c$-approximate reverse $k$-ranks query]    \label{def:query}
Given $\mathbf{U}$, $\mathbf{P}$, a query vector $\mathbf{q}$, $k$, and $c$, this query returns $\mathbf{U}_{c} \subseteq \mathbf{U}$ such that $|\mathbf{U}_{c}| = k$.
Let $\mathbf{u}$ and $\mathbf{u}'$ be the $i$-th user in $\mathbf{U}_{c}$ and $\mathbf{U}_{rr}$, respectively.
We have $r(\mathbf{q},\mathbf{u},\mathbf{P}) \leq c \times r(\mathbf{q},\mathbf{u}',\mathbf{P})$.
\end{definition}


\section{Related Work}  \label{sec:related-work}
\noindent
\textbf{$k$-MIPS works.}
First, tree-based structures for the exact answer were devised in \cite{ram2012maximum}.
However, due to the curse of dimensionality, it is outperformed by linear scan-based algorithms \cite{li2017fexipro,teflioudi2015lemp}.
To accelerate $k$-MIPS efficiency, approximation algorithms were also developed \cite{shrivastava2014asymmetric,liu2020understanding,guo2020accelerating}.
Our problem differs from $k$-MIPS, so the above algorithms are unavailable for our problem.

\vs
\noindent
\textbf{Reverse $k$-MIPS works.}
Simpfer \cite{amagata2021reverse,amagata2023reverse} is an exact algorithm for the reverse $k$-MIPS problem, and SAH \cite{huang2023sah} is the current state-of-the-art approximation algorithm for this problem.
To quickly process a reverse $k$-MIPS query, Simpfer employs linear scan, whereas SAH employs locality-sensitive hashing.

The reverse $k$-ranks problem can be solved by an existing reverse $k$'-MIPS algorithm, but it cannot be fast, for the following two reasons.
First, recall that the result size of a reverse $k$'-MIPS query is not controllable, so it may be less than $k$.
To obtain a $k$-sized result, we need to test multiple (or many) values of $k$', which is clearly computationally expensive.
Furthermore, we need $O(nmd)$ time to process a reverse $k$'-MIPS query, meaning that this approach is worse than the straightforward algorithm.
Second, the existing algorithms function well under the assumption that we know a maximum $k$' in advance.
It is hard to know the maximum value of $k$' that guarantees a $k$-sized result for a reverse $k$-ranks query.

\vs
\noindent
\textbf{Reverse $k$-ranks works.}
QSRP \cite{bian2024qsrp} is the state-of-the-art algorithm for the reverse $k$-ranks problem in high dimensions.
(Some works \cite{zhang2014reverse,dong2017grid,dutta2020radar} consider low dimensions, thus their techniques suffer from the curse of dimensionality.)
To reduce the number of users that need to obtain $r(\mathbf{q},\mathbf{u},\mathbf{P})$, QSRP summarizes the inner products of all pairs of user-item vectors offline. 
If all inner products are maintained, we need ${\rm\Theta}(nm)$ space, and this is infeasible in practice.
Therefore, QSRP maintains some of them.
Given a reverse $k$-ranks query, it filters users that cannot be in the query result based on the summarized inner products.

Nonetheless, QSRP needs to compute the inner products of all pairs of user-item vectors, incurring ${\rm\Omega}(nmd)$ time.
This cost is too large even for pre-processing.
In addition, its time complexity for query processing is still $O(nmd)$, i.e., it is theoretically the same as the straightforward algorithm.

\section{Proposed Algorithm}    \label{sec:proposal}
\noindent
\textbf{Main idea.}
The main computational bottleneck of the $c$-approximate reverse $k$-ranks problem is to compute the rank of $\mathbf{q}$ for each user vector $\mathbf{u}$, see \Cref{def:rank}.
This cost is $O(md)$.
Our idea for avoiding this cost is to obtain a lower-bound rank $r^{\downarrow}_{\mathbf{u}}$ and an upper-bound rank $r^{\uparrow}_{\mathbf{u}}$, i.e., $r^{\downarrow}_{\mathbf{u}} \leq r(\mathbf{q},\mathbf{u},\mathbf{P}) \leq r^{\uparrow}_{\mathbf{u}}$.
Let $R^{\downarrow}_{k}$ ($R^{\uparrow}_{k}$) be the $k$-th smallest lower-bound (upper-bound) rank among all $r^{\downarrow}_{\mathbf{u}}$ ($r^{\uparrow}_{\mathbf{u}}$).
The following observation is directly derived from \Cref{def:query}:

\begin{lemma}   \label{lemma:filter}
(1) If $r^{\uparrow}_{\mathbf{u}} \leq c \times R^{\downarrow}_{k}$, $\mathbf{u}$ can be in $\mathbf{U}_{c}$.
(2) If $r^{\downarrow}_{\mathbf{u}} > R^{\uparrow}_{k}$, $\mathbf{u} \notin \mathbf{U}_{c}$.
\end{lemma}

If $\mathbf{u}$ has one of the above cases, we can skip computing $r(\mathbf{q},\mathbf{u},\mathbf{P})$.
Even if it does not have either case, we can estimate $r(\mathbf{q},\mathbf{u},\mathbf{P})$ from linear interpolation using $r^{\downarrow}_{\mathbf{u}}$ and $r^{\uparrow}_{\mathbf{u}}$.
This again avoids computing $r(\mathbf{q},\mathbf{u},\mathbf{P})$.
In this way, our query processing algorithm can run in $O(nd)$ time.
Now the question is how to obtain $r^{\downarrow}_{\mathbf{u}}$ and $r^{\uparrow}_{\mathbf{u}}$.
The subsequent sections answer this question.

\subsection{Data Structure} \label{sec:proposal:index}
We propose a data structure, \textit{rank table}, to efficiently obtain $r^{\downarrow}_{\mathbf{u}}$ and $r^{\uparrow}_{\mathbf{u}}$.
Consider two inner products $t_{i}$ and $t_{i+1}$ ($t_{i} \leq t_{i+1}$).
If $t_i \leq \mathbf{u} \cdot \mathbf{q} \leq t_{i+1}$, we have
\begin{equation*}
    1 + \sum_{\mathbf{p}\in\mathbf{P}}\mathbb{I}[\mathbf{u}\cdot \mathbf{p} >t_{i+1}] \leq r(\mathbf{u},\mathbf{q},\mathbf{P}) \leq 1 + \sum_{\mathbf{p}\in\mathbf{P}}\mathbb{I}[\mathbf{u}\cdot \mathbf{p} >t_i].
\end{equation*}
From this observation, we prepare multiple inner product thresholds $t_{\mathbf{u}_i,1}, ..., t_{\mathbf{u}_i,\tau}$ for each user $\mathbf{u}_{i}$, where $\tau$ is the number of thresholds, and pre-compute the rank of an item vector $\mathbf{p}$ for $\mathbf{u}_{i}$ when $\mathbf{u}_{i} \cdot \mathbf{p} = t_{i,j}$.
That is, a rank table $\mathcal{T}$ is represented as a matrix, and $\mathcal{T}_{i,j}$ represents the $i$-th row and $j$-th column and is the rank of an item $\mathbf{p}$ for $\mathbf{u}_{i}$ when $\mathbf{u}_{i} \cdot \mathbf{p} = t_{i,j}$.

\begin{figure}[!t]
    \centering
    \includegraphics[width=0.90\linewidth]{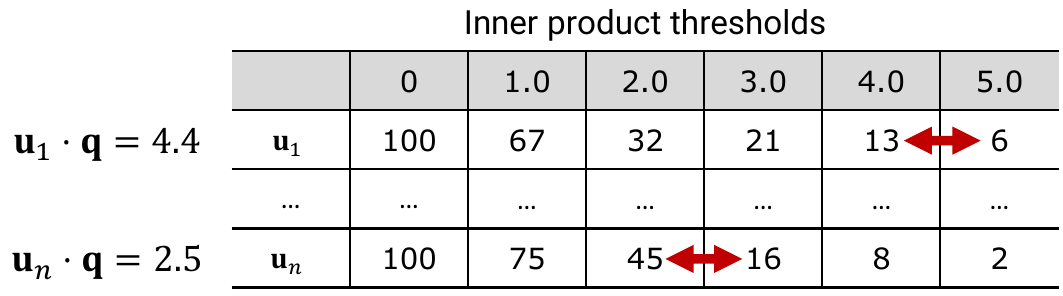}
    \caption{Example of a rank table}
    \label{fig:rank-table}
\end{figure}

\Cref{fig:rank-table} illustrates an example of a rank table.
Assume that $\mathbf{u}_{1} \cdot \mathbf{q} = 4.4$, and we have $6 \leq r(\mathbf{q},\mathbf{u}_1,\mathbf{P}) \leq 13$, so we can obtain $r^{\downarrow}_{\mathbf{u}_1} = 6$ and $r^{\uparrow}_{\mathbf{u}_1} = 13$.

\subsection{Pre-processing Algorithm}
We compute a rank table in a pre-processing step.
This step is done only once, and the rank table is general to any $c$, $\mathbf{q}$, and $k$.
\Cref{algo:pre-processing} describes the pre-processing algorithm.
This algorithm has the following three steps.

\begin{algorithm}[!t]
    \caption{\textsc{Pre-processing}}
    \label{algo:pre-processing}
    \DontPrintSemicolon
    \KwIn {$\mathbf{U}$, $\mathbf{P}$, $\tau$, $\omega$ ($\#$partitions), and $s$ ($\#$samples per partition)}
    \textbf{foreach} $\mathbf{p}_{i} \in \mathbf{P}$ \textbf{do} Compute $\|\mathbf{p}_{i}\|$\;
    Sort $\mathbf{P}$ in descending order of norm\;
    Partition $\mathbf{P}$ into $\omega$ disjoint equally-sized buckets $\mathbf{P}_1, \mathbf{P}_2, \dots, \mathbf{P}_\omega$\;
    $\mathbf{P}_{1}^{sample}, \mathbf{P}_{2}^{sample}, \dots, \mathbf{P}_{\omega}^{sample} \gets \varnothing$\;
    \ForEach {$i \in [1, \omega]$}
    {
        $\mathbf{P}_{i}^{sample} \gets s$ random samples in $\mathbf{P}_{i}$
    }
    $\mathcal{T} \gets n \times \tau$ matrix (all values are initialized by 1)\;
    \ForEach {$\mathbf{u}_{i} \in \mathbf{U}$}
    {
        $T_{\mathbf{u}_{i}} \gets \varnothing$\;
        \textbf{foreach} $j \in [1,\tau]$ \textbf{do} $T_{\mathbf{u}_{i}} \gets T_{\mathbf{u}_{i}} \cup \{f_{\text{min}} + (j - 1) \cdot \frac{f_{\max} - f_{\min}}{\tau - 1}\}$\;
        \ForEach {$l \in [1,\omega]$}
        {
            $A_{\mathbf{u}_{i}} \gets [0, 0, \dots, 0] \quad (|A_{\mathbf{u}_{i}}| = \tau)$\;
            \ForEach {$\mathbf{p} \in \mathbf{P}_{l}^{sample}$}
            {
                \ForEach {$j \in [1,\tau]$}
                {
                    \eIf {$\mathbf{u}_{i} \cdot \mathbf{p} > t_{\mathbf{u}_{i},j}$}
                    {
                        $A_{\mathbf{u}_{i}}[j] \gets A_{\mathbf{u}_{i}}[j] + 1$
                    }{
                        \textbf{break}
                    }
                }
            }
            \textbf{foreach} $j \in [1,\tau]$ \textbf{do} $\mathcal{T}_{i,j} \gets \mathcal{T}_{i,j} + \frac{A_{\mathbf{u}_{i}}[j] \times |\mathbf{P}_{l}|}{s}$
        }
    }
\end{algorithm}

\vs
\noindent
\textbf{(1) Norm computation.}
We compute the Euclidean norm of $\mathbf{p}$, $\|\mathbf{p}\|$, for each $\mathbf{p} \in \mathbf{P}$, and sort the item vectors in descending order of the norm.
For ease of presentation, we assume that $\mathbf{P} = \{\mathbf{p}_1, \mathbf{p}_2, ..., \mathbf{p}_m\}$ and, for each $i \in [1,m-1]$, $\|\mathbf{p}_i\| \geq \|\mathbf{p}_{i+1}\|$.
Then, we partition $\mathbf{P}$ into $w$ disjoint equally-sized subsets $\mathbf{P}_1$, $\mathbf{P}_2$, ..., $\mathbf{P}_{\omega}$.

\vs
\noindent
\textbf{(2) Preparing multiple inner product thresholds.}
Let $T_{\mathbf{u}}$ be a set of inner product thresholds for a user $\mathbf{u}$, and $T_{\mathbf{u}} = \{t_{\mathbf{u},1}, t_{\mathbf{u},2}, \cdots, t_{\mathbf{u},\tau}\}$, where $\tau$ is the number of thresholds (i.e., the number of columns).
Let $f_{\max}(\mathbf{u},\mathbf{P}) = \max_{\mathbf{p} \in \mathbf{P}}\mathbf{u} \cdot \mathbf{p}$.
Also, let $f_{\min}(\mathbf{u},\mathbf{P}) = \min_{\mathbf{p} \in \mathbf{P}}\mathbf{u} \cdot \mathbf{p}$\footnote{Computing $f_{\max}$ and $f_{\min}$ needs $O(md)$ time, but they can respectively be replaced with the maximum and minimum domain values, which yields $O(1)$ time computation.}.
To obtain uniform inner product thresholds in $[f_{\min}, f_{\max}]$, we set $t_{\mathbf{u},i}$ ($i \in [1, \tau]$) as follows:
\begin{equation*}
    t_{\mathbf{u},i} = f_{\min}(\mathbf{u}, \mathbf{P}) + (i - 1) \times \frac{f_{\text{max}}(\mathbf{u}, \mathbf{P}) - f_{\min}(\mathbf{u}, \mathbf{P})}{\tau - 1}.
\end{equation*}

\begin{figure}[!t]
    \begin{center}
        \subfigure[Amazon-K]{%
    	\includegraphics[width=0.48\linewidth]{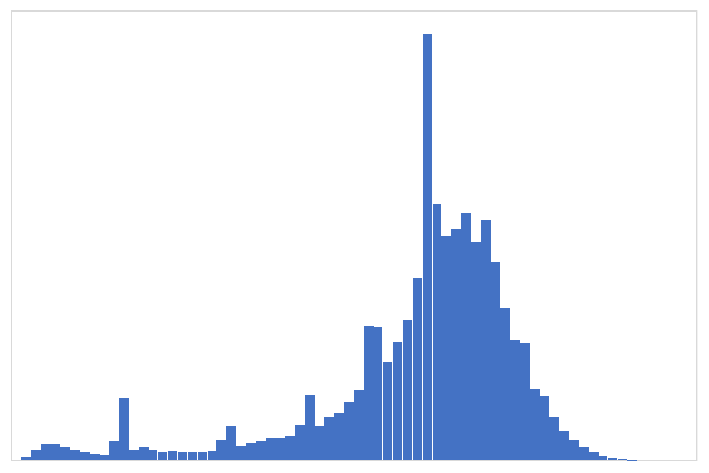}     \label{fig:norm_amazon}}
        \subfigure[MovieLens]{%
    	\includegraphics[width=0.48\linewidth]{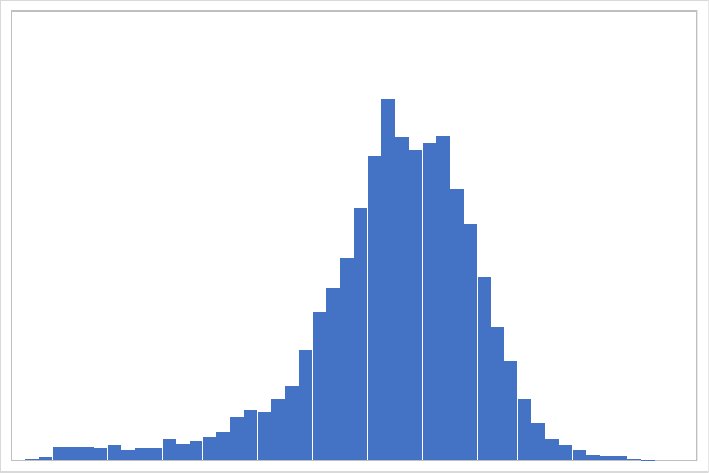}  \label{fig:norm_movielens}}
        \caption{Norm distributions}
        \label{fig:norm}
    \end{center}
\end{figure}

\vs
\noindent
\textbf{(3) Rank table building.}
From the above step, we have
\begin{equation*}
    \mathcal{T}_{i,j} = 1 + \sum_{\mathbf{p} \in \mathbf{P}}\mathbb{I}[\mathbf{u}_i \cdot \mathbf{p} > t_{\mathbf{u}_{i},j}].
\end{equation*}
However, computing $\sum_{\mathbf{p} \in \mathbf{P}}\mathbb{I}[\mathbf{u} \cdot \mathbf{p} > t_{\mathbf{u}_{i},j}]$ requires $O(md)$ time, so doing it for every user incurs $O(nmd)$ time.
To avoid this prohibitive computational cost, we employ cardinality estimation based on random sampling.
Let $\mathbf{P}^{sample}$ be a set of random samples from $\mathbf{P}$.
We can estimate $\mathcal{T}_{i,j}$, and its estimated value $\hat{\mathcal{T}}_{i,j}$ is
\begin{equation*}
    \hat{\mathcal{T}}_{i,j} = 1 + \frac{m \times \sum_{\mathbf{p} \in \mathbf{P}^{sample}}\mathbb{I}[\mathbf{u}_i \cdot \mathbf{p} > t_{\mathbf{u}_{i},j}]}{|\mathbf{P}^{sample}|}
\end{equation*}
It is important to note that, if we use $\mathbf{P}^{sample}$ simply, we suffer from inaccuracy of cardinality estimation.
Because the norm (and inner product) distributions of real-world datasets usually follow a Gaussian distribution (see \Cref{fig:norm}), standard random sampling faces the difficulty of sampling vectors $\mathbf{p}$ such that $\mathbf{u} \cdot \mathbf{p}$ is large and small.

To alleviate this issue, we use the $\omega$ disjoint subsets $\mathbf{P}_1, ..., \mathbf{P}_{\omega}$.
Let $\mathbf{P}^{sample}_{i}$ be a set of random samples in a subset $\mathbf{P}_{i}$.
Also, let $s$ be the sample size.
We have
\begin{equation}
    \hat{\mathcal{T}}_{i,j} = 1 + \sum_{l=1}^{\omega}\frac{|\mathbf{P}_{l}| \times \sum_{\mathbf{p} \in \mathbf{P}_{l}^{sample}}\mathbb{I}[\mathbf{u}_i \cdot \mathbf{p} > t_{\mathbf{u}_{i},j}]}{s}.  \label{eq:cell}
\end{equation}
This computation requires only $O(d)$ time for $\omega = O(1)$ and $s = O(1)$.
Furthermore, due to the random sampling nature, we have $\mathbb{E}[\hat{\mathcal{T}}_{i,j}] = \mathcal{T}_{i,j}$.

\vs
\noindent
\textbf{Space complexity.}
The rank table is an $n \times \tau$ matrix, and $\tau = O(1)$.
Also, the number of inner product thresholds is $n \times \tau$.
Therefore, they consume only $O(n)$ space.

\vs
\noindent
\textbf{Time complexity.}
Step 1 needs ${\rm\Theta}(md)$ time to compute the norm of every item vector.
Also, this step additionally needs $O(m\log m)$ time to sort these item vectors.
In step 2, we need $O(n)$ time to prepare a set of inner product thresholds for every user vector in $\mathbf{U}$.
The main bottleneck of step 3 is to compute \Cref{eq:cell}, and it needs $O(d)$ time for each user vector when $s = O(1)$ and $\omega = O(1)$.
Therefore, \Cref{algo:pre-processing} needs $O((n+m)d + m\log m)$ time.

\subsection{Query Processing Algorithm}
We present our rank table-based algorithm for processing a $c$-approximate reverse $k$-ranks query.
Given $\mathbf{q}$, $k$, and $c$, this algorithm runs the following three steps.

\vs
\noindent
\textbf{(1) Computing lower- and upper-bound ranks.}
First, we compute $\mathbf{u} \cdot \mathbf{q}$ for each $\mathbf{u} \in \mathbf{U}$.
We then compute $r^{\downarrow}_{\mathbf{u}}$ and $r^{\uparrow}_{\mathbf{u}}$.
From the inner product thresholds and rank table, if $t_{\mathbf{u},j} \leq \mathbf{u} \cdot \mathbf{p} \leq t_{\mathbf{u},j+1}$, we have $\hat{\mathcal{T}}_{i,j} \geq r(\mathbf{q},\mathbf{u},\mathbf{P}) \geq \hat{\mathcal{T}}_{i,j+1}$.
Therefore, we set
\begin{equation*}
    r^{\downarrow}_{\mathbf{u}} = \hat{\mathcal{T}}_{i,j+1} \ \text{and} \ r^{\uparrow}_{\mathbf{u}} = \hat{\mathcal{T}}_{i,j}.
\end{equation*}
Then, we compute $R^{\downarrow}_{k}$ ($R^{\uparrow}_{k}$), the $k$-th smallest lower-bound (upper-bound) rank among all $r^{\downarrow}_{\mathbf{u}}$ ($r^{\uparrow}_{\mathbf{u}}$).

\vs
\noindent
\textbf{(2) Filtering user vectors.}
We next filter user vectors with \Cref{lemma:filter}.
If $c \times R^{\downarrow}_{k} \geq R^{\uparrow}_{k}$, it is guaranteed that we have at least $k$ user vectors such that $r(\mathbf{q},\cdot,\mathbf{P}) \leq c \times R^{\downarrow}_{k}$.
Therefore, we insert $k$ user vectors with the smallest $r(\mathbf{q},\cdot,\mathbf{P})$ into $\mathbf{U}_{c}$.
In this case, the search is over.

If $c \times R^{\downarrow}_{k} < R^{\uparrow}_{k}$, we do not have the above guarantee.
We hence insert only user vectors such that $r(\mathbf{q},\cdot,\mathbf{P}) \leq c \times R^{\downarrow}_{k}$ into $\mathbf{U}_{c}$, and filter the user vectors such that $R^{\uparrow}_{k} \leq r(\mathbf{q},\cdot,\mathbf{P})$.
The remaining user vectors are maintained in $\mathbf{U}_{temp}$.

\vs
\noindent
\textbf{(3) Processing not-filtered user vectors.}
If $|\mathbf{U}_{c}| < k$ in the above step, we need to insert $k - |\mathbf{U}_{c}|$ vectors into $\mathbf{U}_{c}$.
We estimate $r(\mathbf{q},\mathbf{u},\mathbf{P})$ with linear interpolation using $r^{\downarrow}_{\mathbf{u}}$ and $r^{\uparrow}_{\mathbf{u}}$ for each $\mathbf{u} \in \mathbf{U}_{temp}$.
We insert $k - |\mathbf{U}_{c}|$ vectors with the smallest estimated $r(\mathbf{q},\cdot,\mathbf{P})$ into $\mathbf{U}_{c}$.

\vs
\noindent
\textbf{Time complexity.}
Step 1 requires $O(nd)$ time since we compute $\mathbf{u} \cdot \mathbf{q}$ for every $\mathbf{u} \in \mathbf{U}$.
Step 2 requires $O(n)$ time since we scan $\mathbf{U}$ once.
In step 3, we have $|\mathbf{U}_{temp}| = O(n)$ in the worst case, so this step also requires $O(n)$ time.
In total, our query processing algorithm needs $O(nd)$ time, which is faster than the time of the state-of-the-art \cite{bian2024qsrp}.

\section{Experiment}    \label{sec:experiment}
This section reports our experimental results.
All experiments were conducted on a Ubuntu 22.04 LTS machine with Intel Core i9-10980XE CPU@3.0GHzx and 128GB RAM.

\vs
\noindent
\textbf{Dataset.}
We used three real world datasets: Amazon-K \cite{he2016ups}, MovieLens\footnote{\url{https://grouplens.org/datasets/movielens/}}, and Netflix\footnote{\url{https://www.cs.uic.edu/liub/Netflix-KDD-Cup-2007.html}}.
Amazon-K is a set of rating data obtained in Amazon Kindle.
The numbers of users and items are respectively 1,406,890 and 430,530.
MovieLens is the MovieLens 25M dataset.
The numbers of users and items are respectively 162,541 and 59,047.
Netflix is a rating dataset used in Netflix Prize.
The numbers of users and items are respectively 480,189 and 17,770.

We used Matrix Factorization \cite{chin2016libmf} to obtain user and item vectors, and we set $d = 200$.

\vs
\noindent
\textbf{Evaluated algorithm.}
We compared our algorithm (denoted by \textsf{Ours}) with \textsf{QSRP} \cite{bian2024qsrp}.
(Recall that \textsf{QSRP} is the state-of-the-art algorithm for reverse k-ranks queries, and \cite{bian2024qsrp} demonstrates that the performance of \textsf{QSRP} is much better than those of existing algorithms for low dimensions.)
We extended \textsf{QSRP} so that it can deal with our problem.
These algorithms were implemented in C++ and compiled by g++ 11.4.0 with -O3 optimization.
We ran pre-processing with 18 threads, and query processing was done with a single thread.
The data structure of \textsf{QSRP} was prepared so that its memory consumption is similar to our data structure for fair comparison.

\vs
\noindent
\textbf{Criteria.}
We used 1,000 random item vectors in $\mathbf{P}$ as queries and measured the following criteria.
\begin{itemize}
    \setlength{\leftskip}{-4.0mm}
    \item   Average running time.
    \item   Average accuracy:
            Given the result of a query, its accuracy is:
            \begin{equation*}
                \text{Accuracy} =\frac{\sum_{k}\mathbb{I}[r(\mathbf{q},\mathbf{u},\mathbf{P}) \leq c \times r(\mathbf{q},\mathbf{u}',\mathbf{P})]}{k},
            \end{equation*}
            where $\mathbf{u} \in \mathbf{U}_{c}$ and $\mathbf{u}' \in \mathbf{U}_{rr}$ are defined in \Cref{def:query}.
            Because of the estimation in \Cref{eq:cell}, the accuracy of \textsf{Ours} is not guaranteed to be 1.
    \item   Average overall ratio:
            This metric measures how $\mathbf{U}_{c}$ is similar to $\mathbf{U}_{rr}$.
            \begin{equation*}
                \text{Overall ratio} =\frac{1}{k}\sum_{k}\frac{r(\mathbf{q},\mathbf{u},\mathbf{P})}{r(\mathbf{q},\mathbf{u}',\mathbf{P})}.
            \end{equation*}
\end{itemize}

\begin{table}[!t]
    \centering
    \caption{Impact of $\tau$}
    \label{tab:tau}
    \begin{tabular}{l|rrr|rrr} \toprule
                    & \multicolumn{3}{c|}{Time [msec]}  & \multicolumn{3}{c}{Overall ratio} \\ \hline
        $\tau$      & 100       & 500     & 1000        & 100   & 500   & 1000              \\ \midrule
        Amazon-K    & 479.27    & 501.87  & 537.12      & 1.01  & 1.01  & 1.01              \\
        MovieLens   & 54.95     & 57.77   & 61.72       & 1.05  & 1.03  & 1.03              \\
        Netflix     & 166.76    & 171.85  & 183.70      & 1.12  & 1.10  & 1.10              \\ \bottomrule
                    & \multicolumn{3}{c|}{Memory usage [GB]}    \\ \hline
        $\tau$      & 100   & 500   & 1000                  \\ \midrule
        Amazon-K    & 2.09  & 4.23  & 6.92                  \\
        MovieLens   & 0.25  & 0.50  & 0.81                  \\ 
        Netflix     & 0.61  & 1.34  & 2.26                  \\
        \bottomrule
    \end{tabular}
\end{table}

\subsection{Tuning $\tau$}
We first tune $\tau$ (the number of columns in the rank table) for \textsf{Ours}.
\Cref{tab:tau} shows that, as $\tau$ increases, the query processing time and memory usage increase, but the overall ratio almost does not vary.
(We omit accuracy here because of space limitation, and it follows the same tendency as the overall ratio.)
From this result, we set $\tau = 500$ in the subsequent experiments.

\subsection{Pre-processing Evaluation}
We measured the pre-processing times of \textsf{QSRP} and \textsf{Ours}.
\Cref{tab:offline} shows the result.
Although \textsf{Ours} terminates within a reasonable time, \textsf{QSRP} needs much longer time than \textsf{Ours}.
In particular, \textsf{QSRP} took a significantly long time (about five hours) for a large dataset (i.e., Amazon-K).
This long delay limits the practical usage of \textsf{QSRP}, whereas \textsf{Ours} does not have this issue.

\begin{table}[!t]
    \centering
    \caption{Pre-processing time [sec]}
    \label{tab:offline}
    \begin{tabular}{crrr} \toprule
                        & Amazon-K   & MovieLens     & Netflix   \\ \midrule
        \textsf{Ours}   & 219.46     & 24.22         & 69.21     \\
        \textsf{QSRP}   & 17924.85   & 242.43        & 148.60    \\ \bottomrule
    \end{tabular}
\end{table}
\begin{figure}[!t]
    \begin{center}
        \subfigure[Time (Amazon-K)]{%
    	\includegraphics[width=0.48\linewidth]{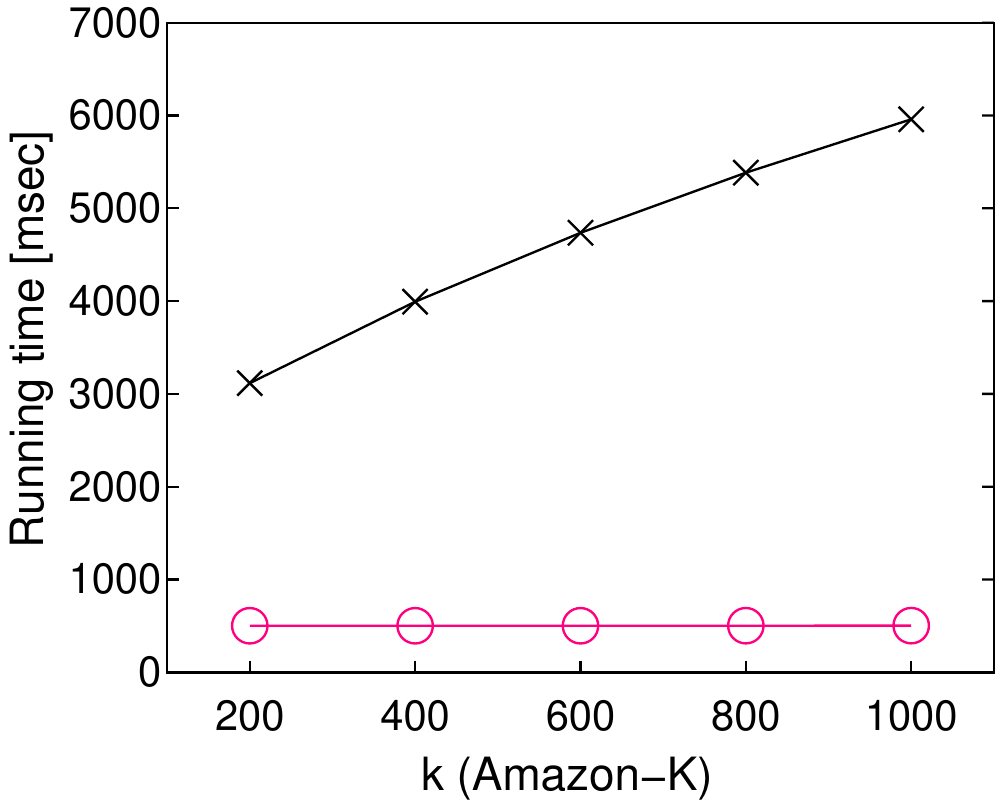}      \label{fig:amazonk-k_time}}
        \subfigure[Accuracy (Amazon-K)]{%
    	\includegraphics[width=0.48\linewidth]{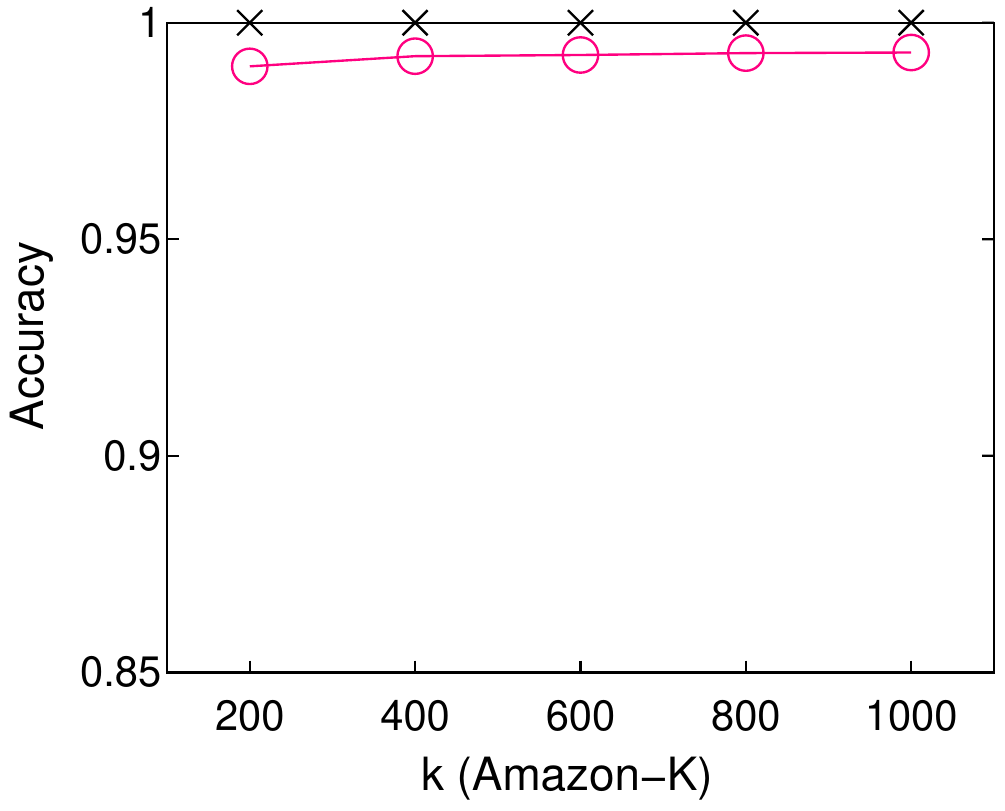}  \label{fig:amazonk-k_accuracy}}
        \subfigure[Overall ratio (Amazon-K)]{%
            \includegraphics[width=0.48\linewidth]{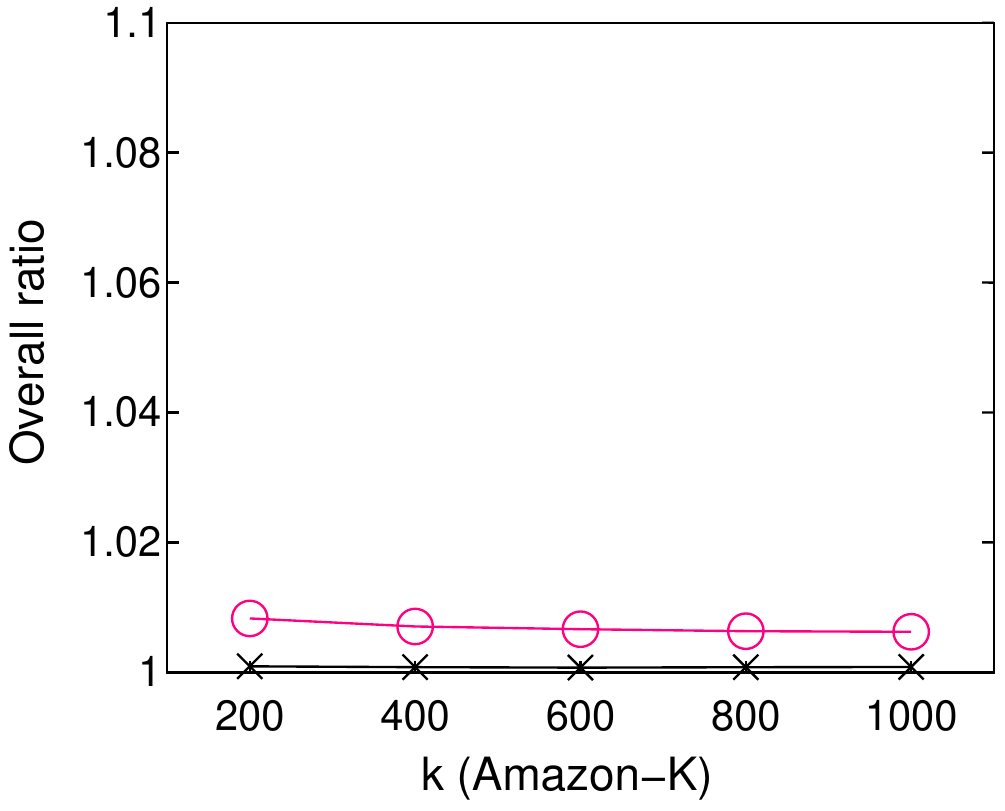}     \label{fig:amazonk-k_ratio}}
        \subfigure[Time (MovieLens)]{%
    	\includegraphics[width=0.48\linewidth]{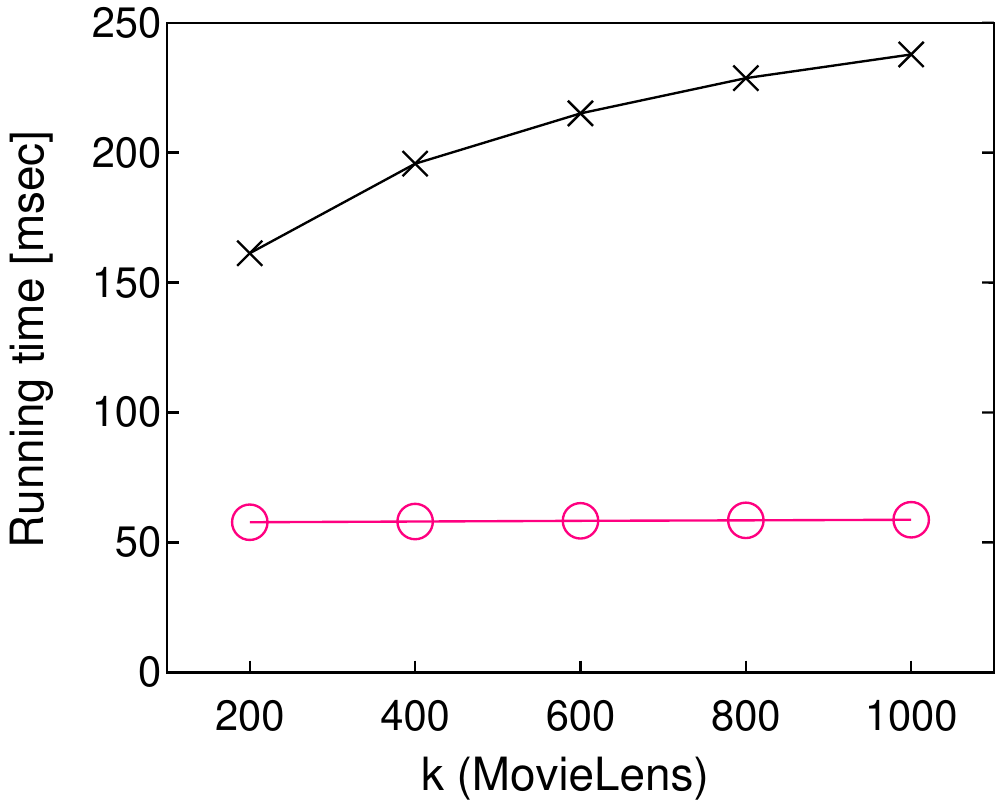}      \label{fig:movielens-k_time}}
        \subfigure[Accuracy (MovieLens)]{%
    	\includegraphics[width=0.48\linewidth]{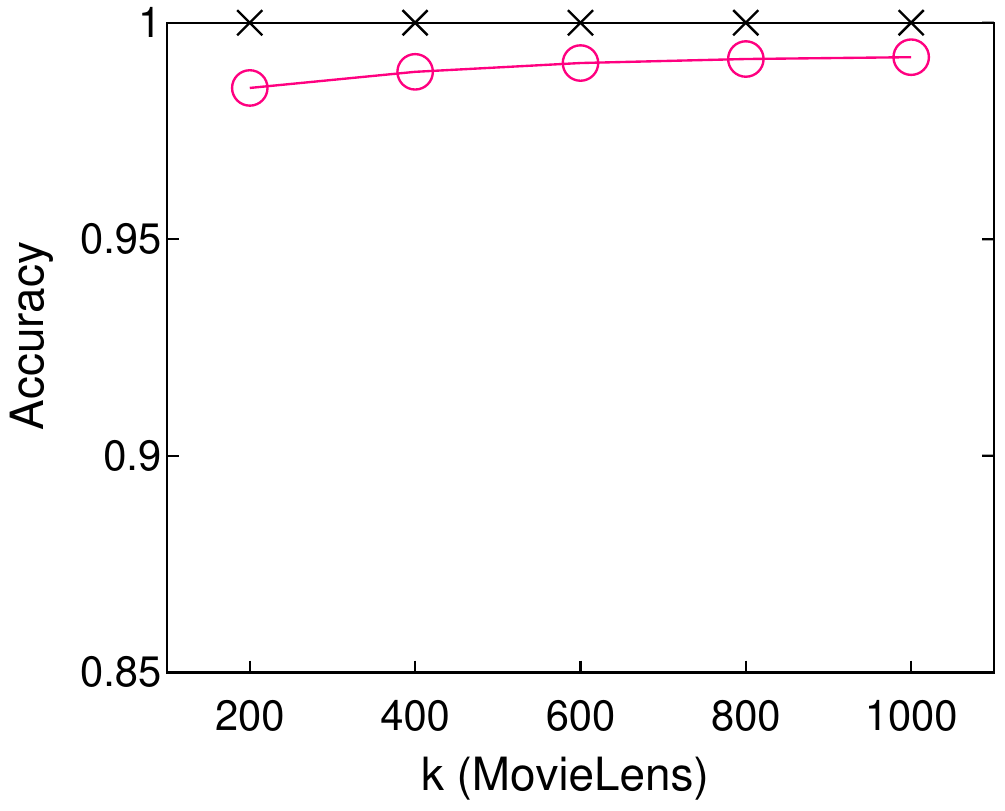}  \label{fig:movielens-k_accuracy}}
        \subfigure[Overall ratio (MovieLens)]{%
            \includegraphics[width=0.48\linewidth]{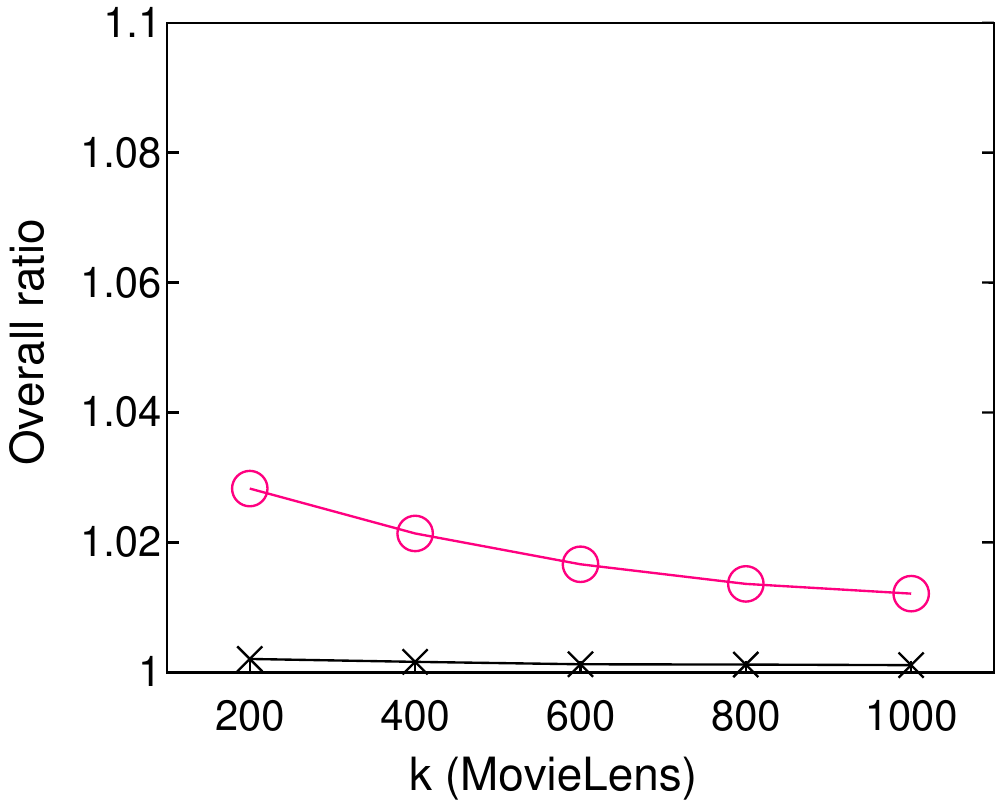}     \label{fig:movielens-k_ratio}}
        \subfigure[Time (Netflix)]{%
    	\includegraphics[width=0.48\linewidth]{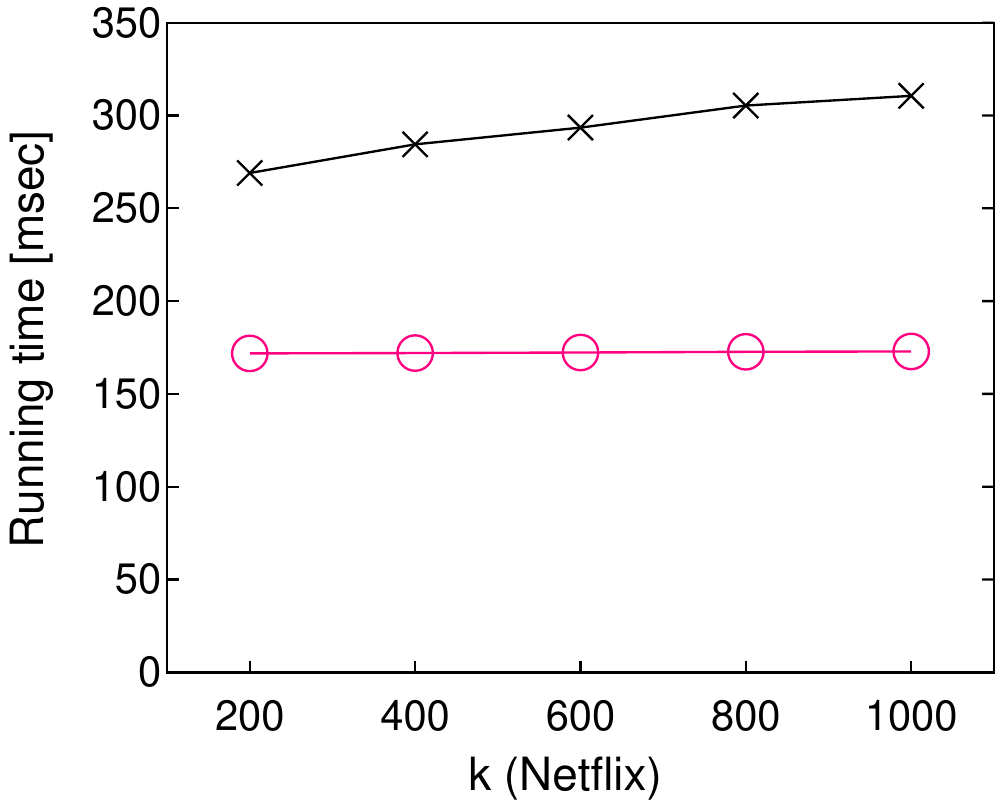}      \label{fig:netflix-k_time}}
        \subfigure[Accuracy (Netflix)]{%
    	\includegraphics[width=0.48\linewidth]{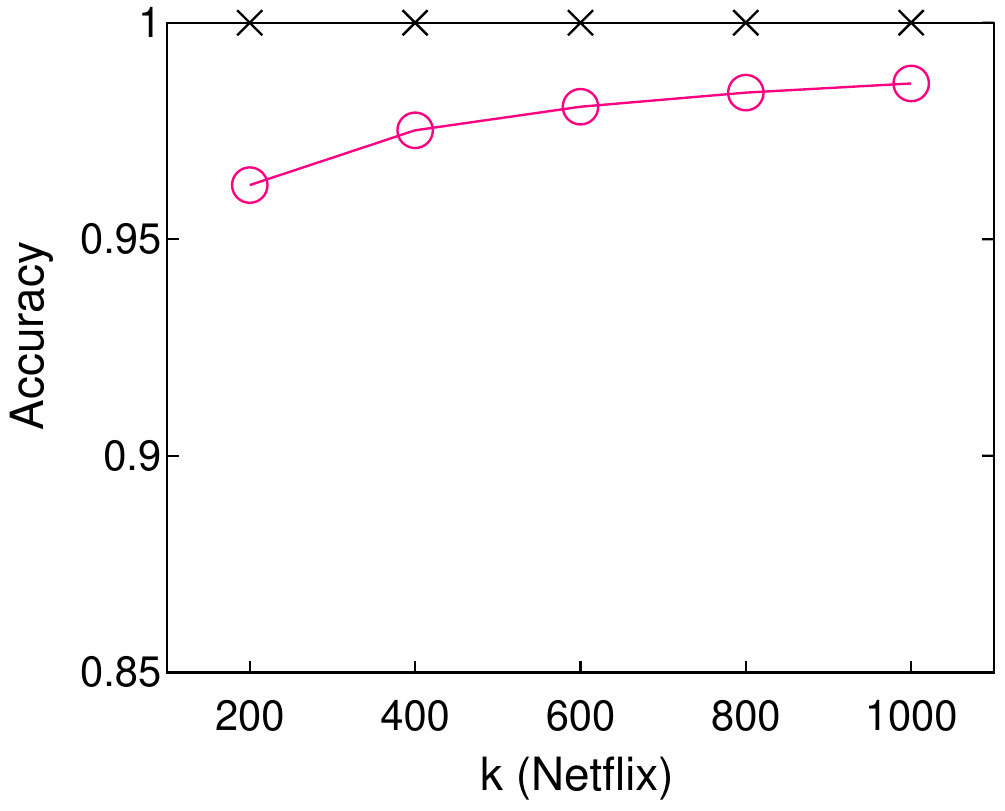}  \label{fig:netflix-k_accuracy}}
        \subfigure[Overall ratio (Netflix)]{%
            \includegraphics[width=0.475\linewidth]{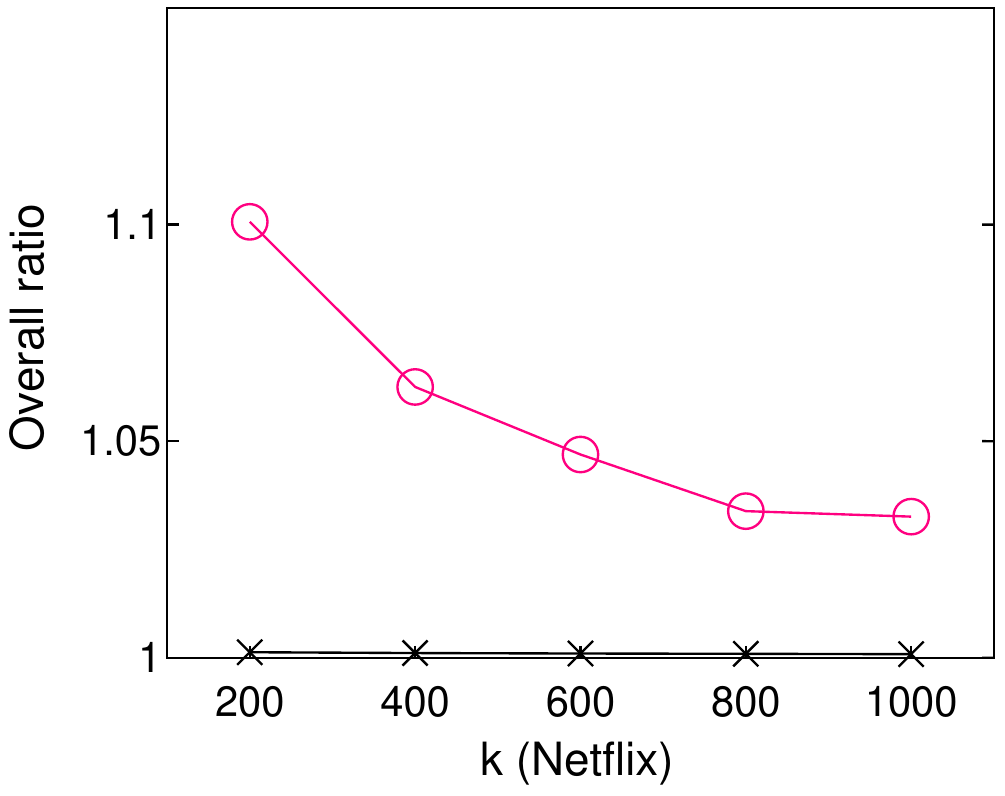}     \label{fig:netflix-k_ratio}}
        \caption{Impact of $k$: ``$\times$'' shows \textsf{QSRP} and ``\textcolor{magenta}{$\circ$}'' shows \textcolor{magenta}{\textsf{Ours}}.}
        \label{fig:k}
    \end{center}
\end{figure}

\subsection{Query Processing Evaluation}
\noindent
\textbf{Impact of $k$.}
We investigate the impact of $k$ on the query processing performance.
\Cref{fig:k} depicts the experimental results.
First, we see that \textsf{Ours} is much (up to 12x) faster than \textsf{QSRP}.
The running time of \textsf{Ours} is not affected by $k$, whereas that of \textsf{QSRP} increases as $k$ increases.

Next, the accuracy of \textsf{Ours} is almost perfect, and its overall ratio is almost 1.
(\textsf{QSRP} guarantees that $r(\mathbf{q},\mathbf{u},\mathbf{P}) \leq c \times r(\mathbf{q},\mathbf{u}',\mathbf{P})$ for each $\mathbf{u} \in \mathbf{U}_c$, so its accuracy is always 1.)
This result demonstrates that \textsf{Ours} practically yields a result similar to $\mathbf{U}_{rr}$ (see \Cref{sec:preliminary}).

\begin{figure}[!t]
    \begin{center}
        \subfigure[Time (Amazon-K)]{%
    	\includegraphics[width=0.48\linewidth]{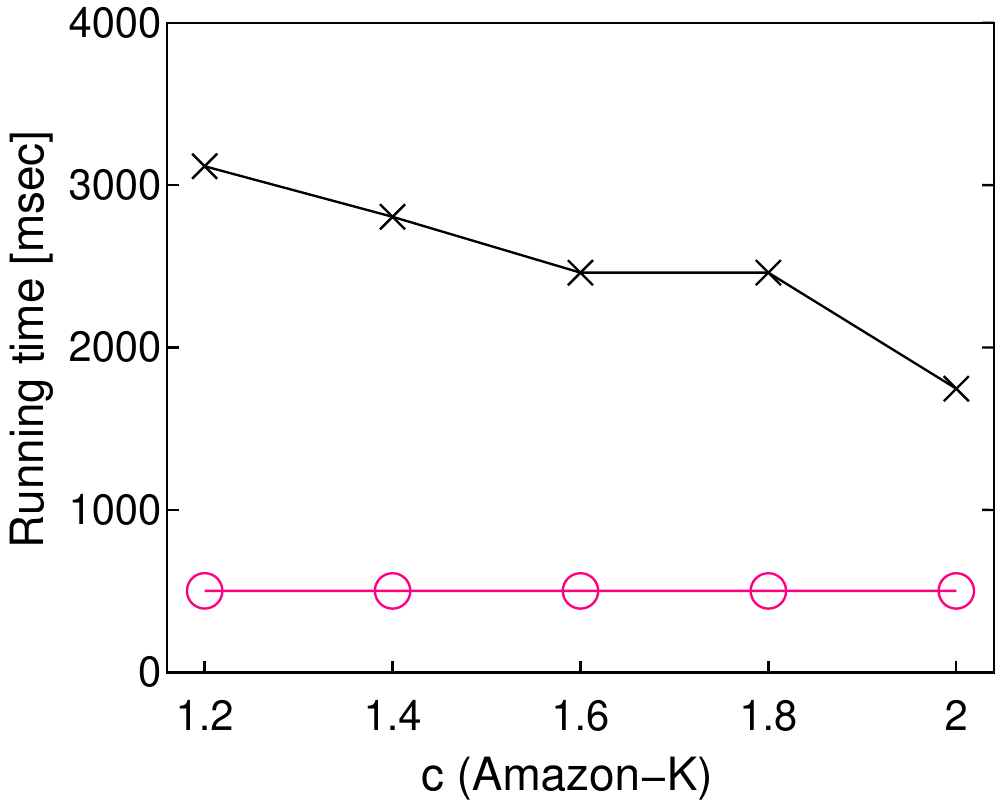}      \label{fig:amazonk-c_time}}
        \subfigure[Accuracy (Amazon-K)]{%
    	\includegraphics[width=0.48\linewidth]{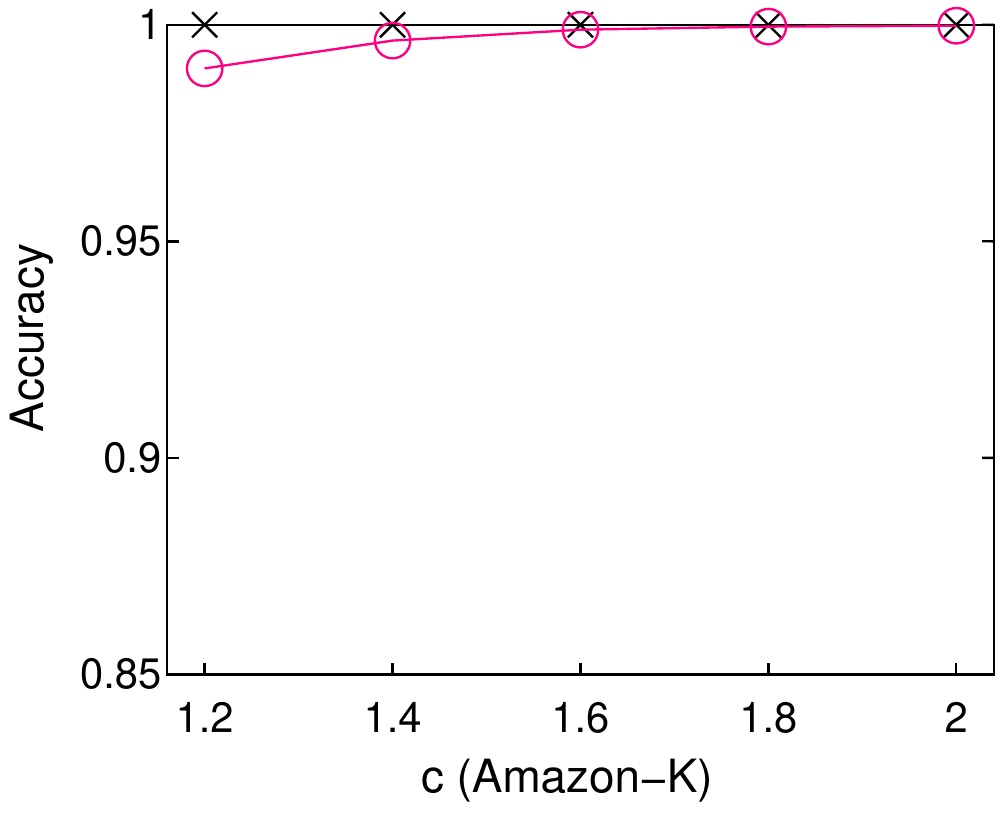}  \label{fig:amazonk-c_accuracy}}
        \subfigure[Overall ratio (Amazon-K)]{%
            \includegraphics[width=0.48\linewidth]{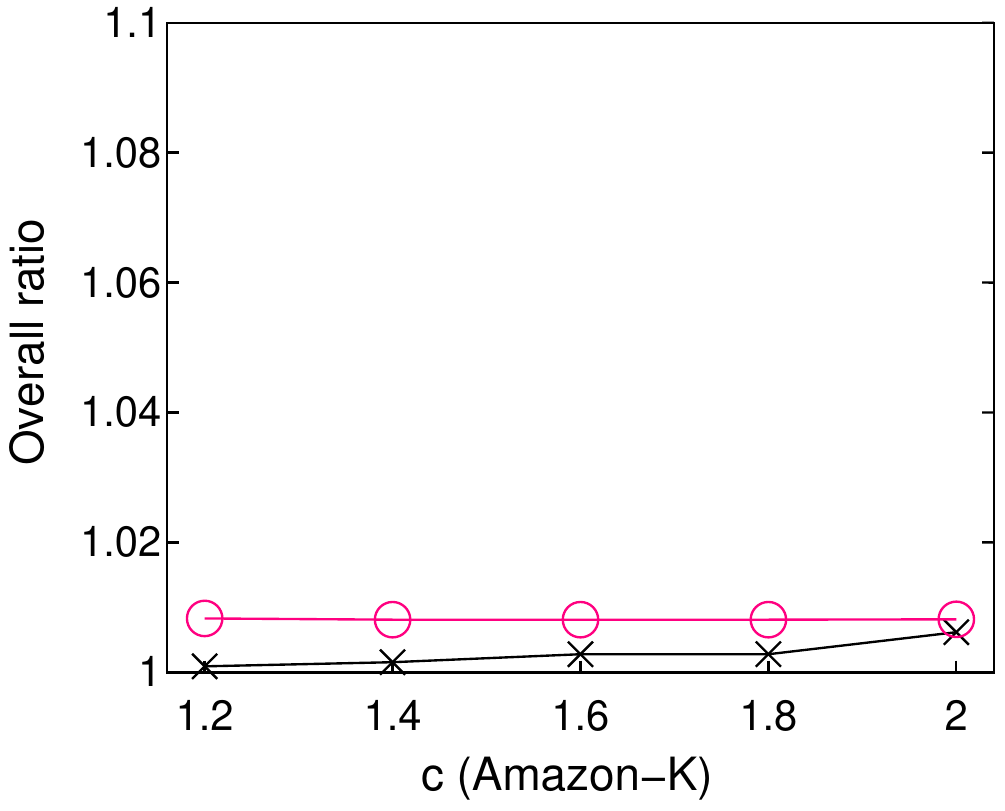}     \label{fig:amazonk-c_ratio}}
        \subfigure[Time (MovieLens)]{%
    	\includegraphics[width=0.48\linewidth]{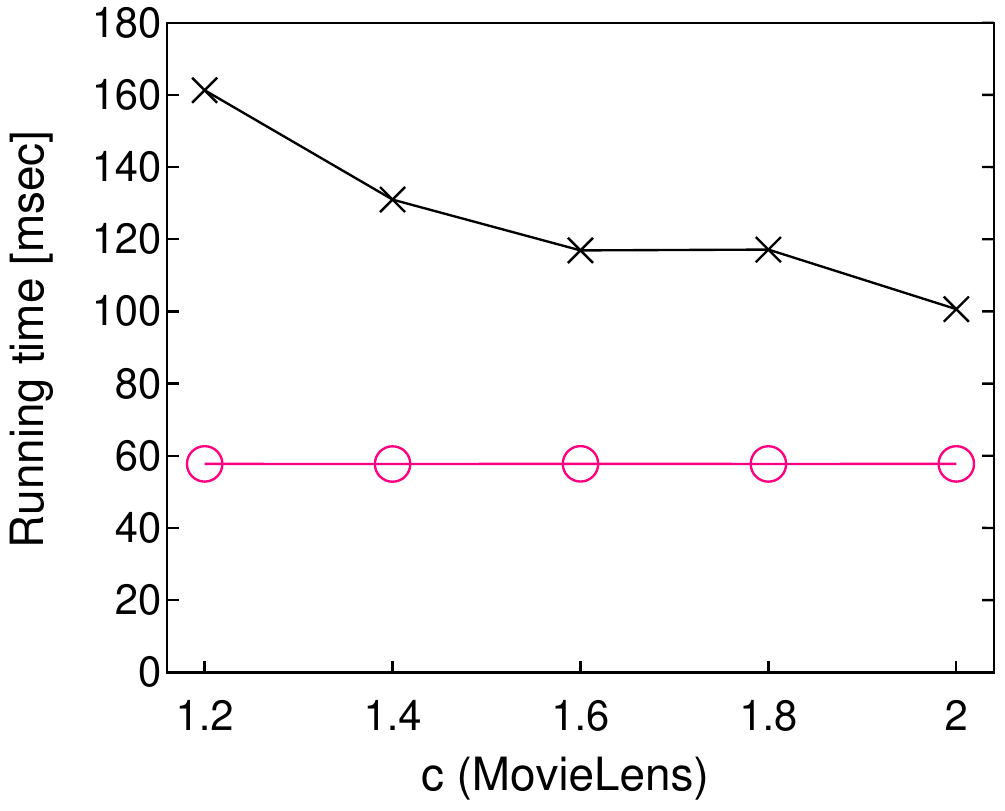}      \label{fig:movielens-c_time}}
        \subfigure[Accuracy (MovieLens)]{%
    	\includegraphics[width=0.48\linewidth]{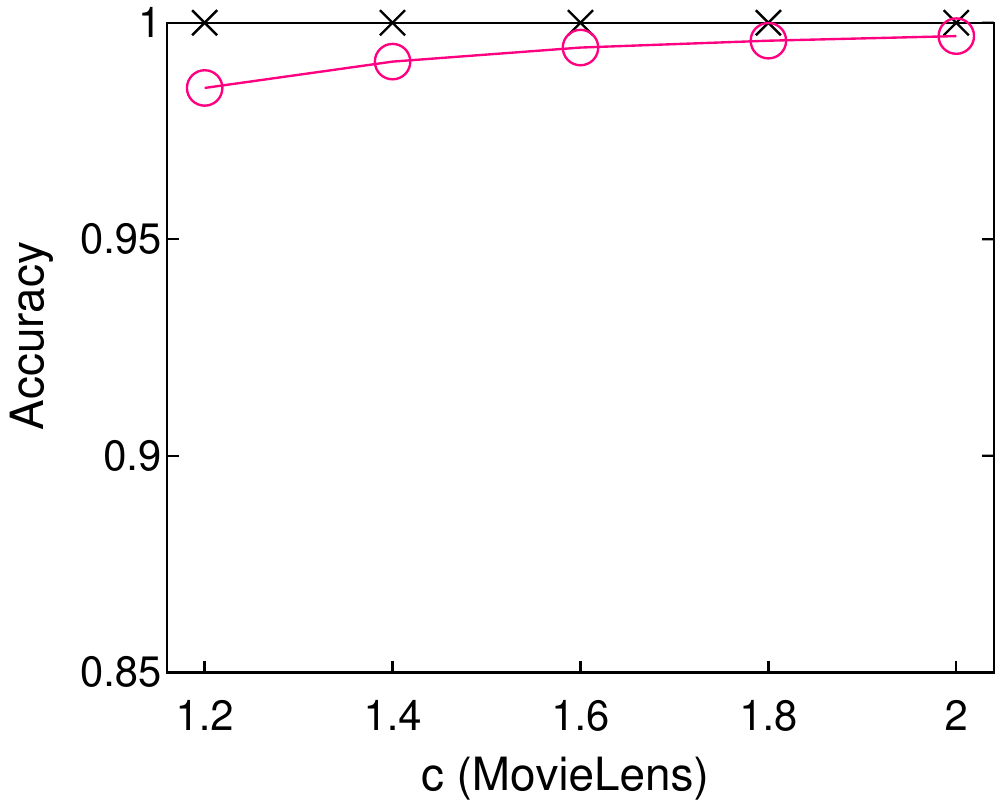}  \label{fig:movielens-c_accuracy}}
        \subfigure[Overall ratio (MovieLens)]{%
            \includegraphics[width=0.48\linewidth]{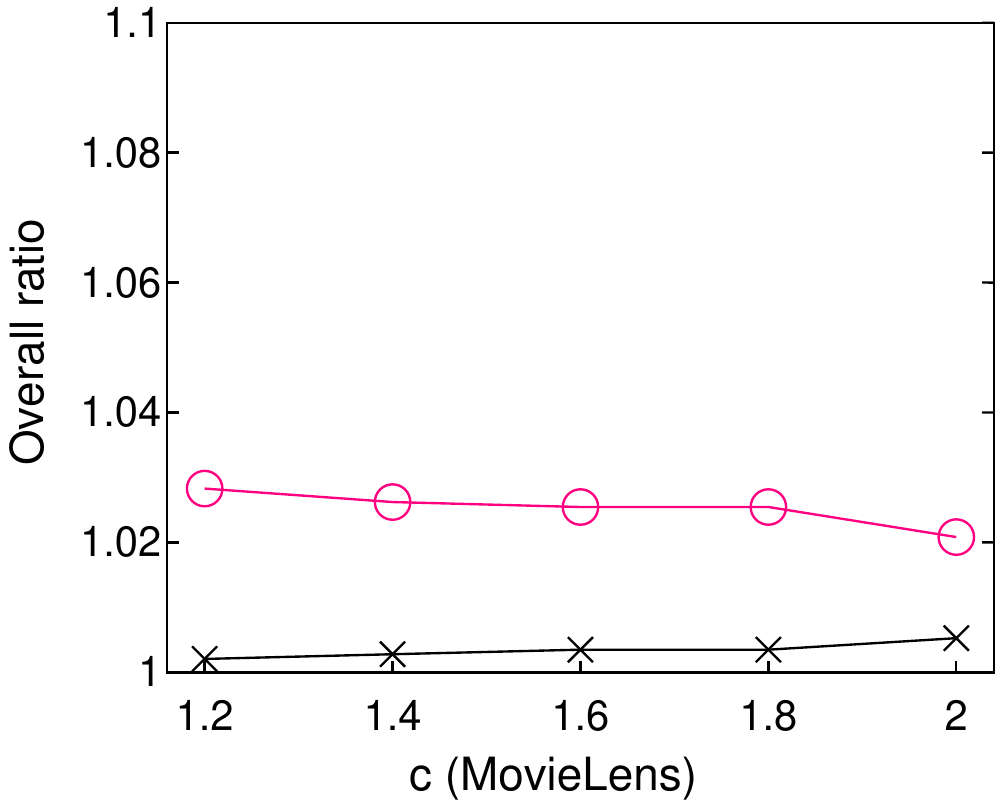}     \label{fig:movielens-c_ratio}}
        \subfigure[Time (Netflix)]{%
    	\includegraphics[width=0.48\linewidth]{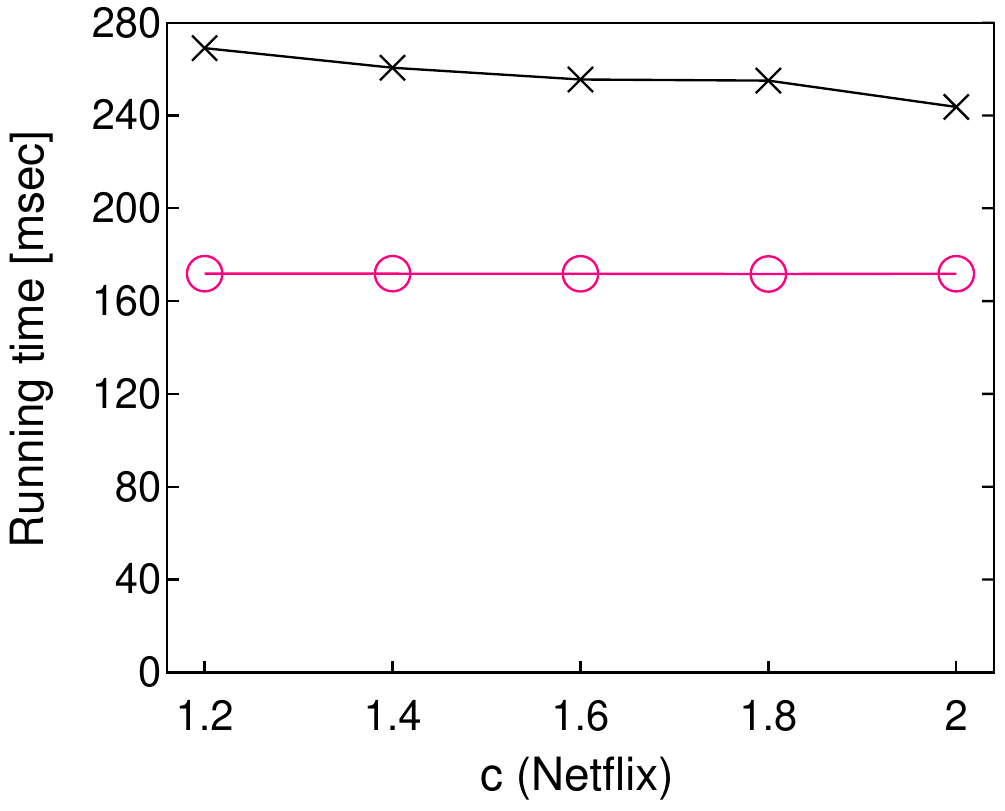}      \label{fig:netflix-c_time}}
        \subfigure[Accuracy (Netflix)]{%
    	\includegraphics[width=0.48\linewidth]{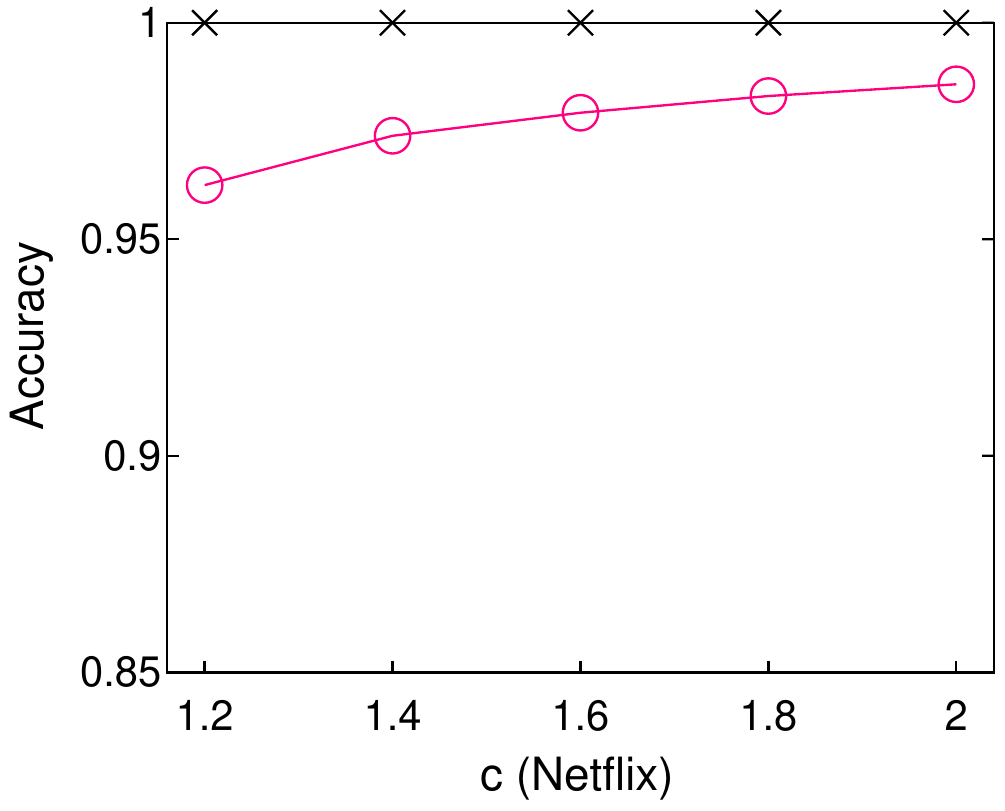}  \label{fig:netflix-c_accuracy}}
        \subfigure[Overall ratio (Netflix)]{%
            \includegraphics[width=0.48\linewidth]{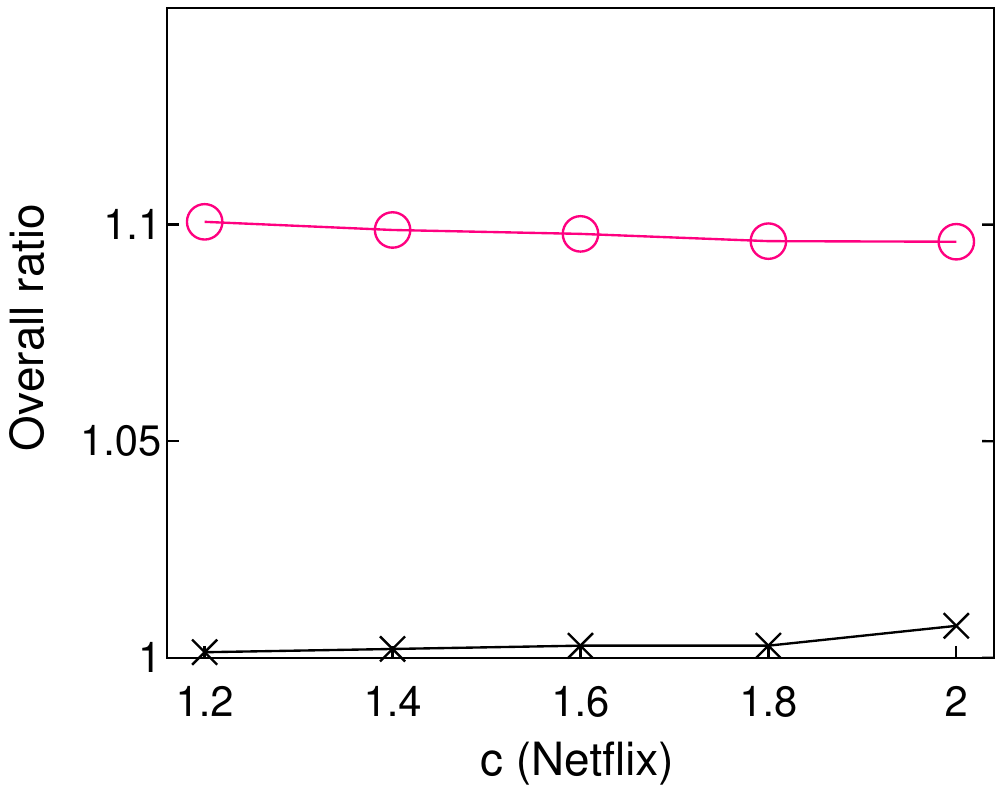}     \label{fig:netflix-c_ratio}}
        \caption{Impact of $c$: ``$\times$'' shows \textsf{QSRP} and ``\textcolor{magenta}{$\circ$}'' shows \textcolor{magenta}{\textsf{Ours}}.}
        \label{fig:c}
    \end{center}
\end{figure}

\vs
\noindent
\textbf{Impact of $c$.}
We next study the impact of $c$, and \Cref{fig:c} shows the result.
Again, \textsf{Ours} is always faster than $\textsf{QSRP}$, and the time of \textsf{Ours} is not affected by $c$.
To understand this observation, \Cref{tab:decomposed} shows the time of each step in \textsf{Ours}.
We see that steps 2 and 3 take negligible time, and step 1 dominates the running time.
Because step 1 is not affected by $k$ and $c$, we have this observation.

As for accuracy and overall ratio, the results of \textsf{Ours} are similar to those in \Cref{fig:k}.
The difference between \textsf{Ours} and \textsf{QSRP} is very slight.
Moreover, the accuracy and overall ratio of \textsf{Ours} is almost 1.

\begin{table}[t]
    \centering
    \caption{Decomposed time of \textsf{Ours} [msec]}
    \label{tab:decomposed}
    \begin{tabular}{lccc} \toprule
                    & Step 1                & Step 2                & Step 3                 \\ \midrule
        Amazon-K    & $5.02 \times 10^2$    & $1.92 \times 10^{-2}$ & $1.98 \times 10^{-4}$  \\
        Movielens   & $5.78 \times 10^1$    & $3.72 \times 10^{-3}$ & $1.07 \times 10^{-4}$  \\
        Netflix     & $1.72 \times 10^2$    & $4.17 \times 10^{-2}$ & $1.24 \times 10^{-4}$  \\
        \bottomrule
    \end{tabular}
\end{table}

\section{Conclusion}    \label{sec:conclusion}
This paper proposed the $c$-approximate reverse $k$-ranks problem and a new algorithm for this problem.
Existing techniques are inefficient for this problem, and even the state-of-the-art algorithm for the reverse $k$-rank problem in high dimensions incurs $O(nmd)$ time.
We overcame this challenge, and our algorithm needs only $O(nd)$ time and yields a much faster time than the state-of-the-art in practice.
Our experimental results demonstrate the efficiency and effectiveness of our algorithm.

\begin{acks}
This work was partially supported by AIP Acceleration Research JPMJCR23U2.
\end{acks}

\bibliographystyle{ACM-Reference-Format}
\bibliography{sigproc}

\end{document}